**Visual Question Answering in Ophthalmology: A Progressive and Practical Perspective**

**Authors:**

Xiaolan Chen, MD[1,#], Ruoyu Chen, MD[1,#], Pusheng Xu, MD[1,#], Weiyi Zhang, MS[1], Xianwen Shang, PhD[1], Mingguang He, MD, PhD[1,2,3], Danli Shi, MD, PhD[1,2]

**Affiliations:**

1. School of Optometry, The Hong Kong Polytechnic University, Kowloon, Hong Kong.
2. Research Centre for SHARP Vision (RCSV), The Hong Kong Polytechnic University, Kowloon, Hong Kong.
3. Centre for Eye and Vision Research (CEVR), 17W Hong Kong Science Park, Hong Kong.

# Contributed equally

**Correspondence:**

\***Prof. Mingguang He**, MD, PhD., Chair Professor of Experimental Ophthalmology, School of Optometry, The Hong Kong Polytechnic University, Kowloon, Hong Kong SAR, China.

Email: mingguang.he@polyu.edu.hk

ORCID records: https://orcid.org/0000-0002-6912-2810

\***Dr Danli Shi**, MD, PhD., The Hong Kong Polytechnic University, Kowloon, Hong Kong SAR, China.

Email: danli.shi@polyu.edu.hk

ORCID records: https://orcid.org/0000-0001-6094-137X

**Abstract**

Accurate diagnosis of ophthalmic diseases relies heavily on the interpretation of multimodal ophthalmic images, a process often time-consuming and expertise-dependent. Visual Question Answering (VQA) presents a potential interdisciplinary solution by merging computer vision and natural language processing to comprehend and respond to queries about medical images. This review article explores the recent advancements and future prospects of VQA in ophthalmology from both theoretical and practical perspectives, aiming to provide eye care professionals with a deeper understanding and tools for leveraging the underlying models. Additionally, we discuss the promising trend of large language models (LLM) in enhancing various components of the VQA framework to adapt to multimodal ophthalmic tasks. Despite the promising outlook, ophthalmic VQA still faces several challenges, including the scarcity of annotated multimodal image datasets, the necessity of comprehensive and unified evaluation methods, and the obstacles to achieving effective real-world applications. This article highlights these challenges and clarifies future directions for advancing ophthalmic VQA with LLMs. The development of LLM-based ophthalmic VQA systems calls for collaborative efforts between medical professionals and AI experts to overcome existing obstacles and advance the diagnosis and care of eye diseases.

**Keywords:** Ophthalmic Visual Question Answering, Large Language Models, Multimodal Image Interpretation, Report Generation, Generative Artificial Intelligence

**Introduction**

Accurate diagnosis of ophthalmic diseases often relies on the comprehensive analysis of multimodal ophthalmic images, including color fundus photographs (CFP), optical coherence tomography (OCT), fundus fluorescein angiography (FFA), scanning laser ophthalmoscopy (SLO), anterior segment photographs and corneal topography, etc. However, interpreting ocular imaging information is time-consuming and experience-dependent in such an image-extensive specialty. Visual question answering (VQA) is an emerging interdisciplinary field that combines computer vision (CV) with natural language processing (NLP). In the past few years, VQA has been introduced in medicine and attracted extensive attention from researchers.[1] Medical visual question answering system is expected to understand free-form questions and generate corresponding answers related to the medical images after image feature extraction, question feature extraction, and feature fusion.[2]

While VQA systems have made certain progress in the biomedical field,[3] research in ophthalmology is still in its early stages, limited to creating essential image captions and reports.[4 5] Despite this, the field shows great potential, suggesting broad future applications. VQA systems can be integrated into basic ophthalmology education, assisting laypeople, or inexperienced young trainees in identifying and understanding clinical images, and could serve as a knowledge assistant for human doctors, providing image-assisted information to improve diagnostic efficiency and reduce misdiagnoses.[6 7] Moreover, for patients, a well-developed ophthalmic VQA system could offer instant professional feedback, reducing frequent hospital visits, thereby alleviating financial burdens economically and lessening the pressure on public healthcare systems.

Despite the significant potential of ophthalmic VQA systems,we still face several challenges during their development.[8] These include the need for accurately annotated multimodal ophthalmic image datasets to train models, the design of advanced frameworks and algorithms to understand and process clinical questions, and the importance of achieving harmonious interaction between artificial intelligence and human doctors.

Notably, large language models (LLMs) have recently been introduced in medical field.[9] LLMs, such as MetaAI's Llama and OpenAI's GPT-4, have shown promising performance in question answering of eye conditions.[10 11] Integrating LLMs into the VQA framework presents new possibilities for database construction, enhanced image understanding, and output optimization, marking a significant leap toward more intelligent VQA systems.[12]

This paper aims to provide the theoretical basis and potential applications of establishing LLM-based VQA systems in ophthalmology. We begin with an overview of key technologies of VQA systems, then summarize the latest progress and potential applications of VQA in the medical field, particularly in ophthalmology. Additionally, we propose potential strategies for optimizing the VQA framework with advanced technologies like LLMs. Finally, we discuss the challenges in developing LLM-based ophthalmic VQA models and their future directions.

**Overview of VQA**

VQA systems synergize advancements in image recognition, NLP, and machine learning to interpret visual data and answer related questions. This section will briefly introduce the core architecture of VQA models, understanding how they process and analyze visual inputs in conjunction with textual queries.

A VQA pipeline generally consists of four parts: an image feature extractor for target images, a textual feature extractor for the query questions, a feature fusion module, and a prediction head for generating answers according to the input image and text (**Figure 1a**).[13]

The image feature extractor is the first critical component of a VQA system, primarily tasked with extracting useful visual features from the input images. Typically, deep convolutional neural networks (CNNs) are widely employed for this purpose, such as ResNet, VGGNet, DenseNet, etc.[14-16] CNNs are capable of extracting hierarchical features from images, from low-level edges and textures to high-level semantic information, providing detailed visual information for further analysis.[17] Pretrained CNN models, such as those pretrained on the ImageNet dataset,[18] are also commonly utilized, with the models undergoing fine-tuning or feature extraction through transfer learning. In addition to CNNs, image feature extraction models based on the Transformer have garnered significant attention. For instance, Yakoub et al.[13] employed the Vision Transformer (ViT), which divides images into patches for embedding and globally models relationships within these images, demonstrating remarkable accuracy in pathological VQA tasks. Other Transformer-based models, such as DeiT and DETR, provide distinct capabilities. DeiT incorporates a distillation token to enhance the performance of the student network,[19] while DETR enables end-to-end object detection.[20] Furthermore, practical practice in medicine often involves comparing current and previous images of the same patient to track disease progression. Image difference captioning techniques appear promising for enhancing the clinical performance of VQA in this aspect.[21 22]

For text feature extraction, previous studies in general domains often apply recurrent neural networks (RNN) and long short-term memory (LSTM). RNNs leverage the cyclic flow of information in sequential data,[23] while LSTM,[24] with its inherent gating mechanism, addresses long-term information dependencies and enables better control of information flow. In recent years, Transformer models, such as BERT and the GPT family, have gained popularity due to their ability to handle long text sequences and parallel computation effectively. BERT is pretrained on a massive text corpus using extensive unsupervised learning, allowing for effective text comprehension and representation.[25] It also indirectly contributes to the text generation process through specific task architectures. An example is the BLIP model, which utilizes BERT for text feature extraction.[26] On the other hand, GPT excels in natural language processing, generating fluent and consistent text outputs based on given text inputs.[27] For instance, the LLaVA model employs GPT (Llama) to tackle more complex vision-language tasks.[28]

For feature fusion, attention mechanism and multimodal pooling technique were widely used in fusing visual and text features in VQA tasks.[3] Multimodal pooling is a common method that reduces data complexity, creating a global and abstract feature representation through weighted averages and maximization calculations.[29 30] Attention mechanism assigns different weights to different input features, allowing the model to "focus" or "concentrate" on the most important or relevant information.[31] Specific implementation methods include employing Dual-Attention to pay equal attention to the image and text,[32] or the use of Dynamic Fusion with Intra- and Inter-modality Attention Flow (DFAF) to robustly capture the high-level interactions between language and vision domains.[33]

The prediction head serves as the final part of the VQA system, generating the ultimate answer based on joint representation. For binary or multiple-choice questions, the system may utilize a multi-layer perceptron for classification tasks.[34] For open-ended questions, generative models, such as GPT, are the primary choices. These models are capable of generating grammatically correct and fluent natural language outputs.[5] The recent progress and integration of LLMs have greatly enhanced the generation of language, resulting in answers that closely align with the image content and adhere to human linguistic patterns.[35]

**Current Progress and Potential Uses of VQA in Ophthalmology**

*Ophthalmic VQA datasets*

Currently, most existing ophthalmological datasets consist of eye images with brief labels, and there is a scarcity of comprehensive ophthalmic VQA datasets. The DME VQA dataset, derived from the IDRiD and eOphta datasets, encompasses both healthy and unhealthy fundus images. Each image is accompanied by a set of predefined questions, including inquiries about specific regions (e.g., "Are there hard exudates in this region?"), along with corresponding masks indicating the region's location.[36] OphthalVQA represents an initial effort to develop a multimodal VQA dataset.[11] It comprises 600 image-QA pairs across six imaging modalities: slit lamp, SLO, CFP, OCT, FFA, and ultrasound images. These VQA pairs cover a wide range of common questions related to modality recognition, disease diagnosis, examination, and treatment, thus serving as a valuable resource for training and evaluating ophthalmic VQA models.

*Advancements in Ophthalmic Visual Question Answering*

**Closed-form VQA**: Closed-form VQA tasks are commonly regarded as classification tasks that aim to provide a limited number of answers. These answers are typically presented as predetermined options, such as a short list of multiple choices, yes/no responses, or numerical ratings. For instance, Tascon et al.[36] incorporated question relationships into their VQA model training and evaluated the model's performance in grading diabetic macular edema using fundus images. The ChatFFA model fine-tuned the BLIP framework with FFA images to answer multiple-choice and true/false questions, demonstrating strong performance in both automated and human evaluations.[37]

**Free-form VQA:** Free-form VQA tasks facilitate responses in the form of phrases or sentences, enabling detailed and fine-grained answers without being restricted by predefined options. Early research on image captioning and report generation laid the groundwork for generating free-text descriptions. Initially, technologies such as ResNet-LSTM, CNN-BLSTM models, and Expert-defined Keywords were utilized to generate preliminary descriptive captions for ophthalmic images, enhancing the efficiency of diagnosis and disease management.[38-41] As NLP progressed, more complex VQA systems were able to automatically generate comprehensive reports from ophthalmic images,[42-44] thereby improving the efficiency of medical report writing for healthcare professionals.

However, image captioning or report generation models can only generate fixed formats of image-text QA and fail to fulfill the requirements for complex question answering in smart clinics. Researchers began exploring the integration of LLMs to construct more human-like QA dialogue systems. OphGLM constructed a pipeline for diagnosing common ophthalmic diseases and lesion segmentation based on fundus images, successfully developing an ophthalmic-specific vision-language assistant by integrating visual capabilities into LLMs.[45] Additionally, FFA-GPT and ICGA-GPT systems, which combine image-to-text conversion models with LLMs to provide interactive image-based question-answering for FFA and ICGA images through a two-step approach.[12 46] Another study leveraged a large number of image-QA pairs to develop a true bilingual VQA model based on FFA (ChatFFA), which has shown great potential in various tasks.[37]

*Potential Application of VQA in Ophthalmology*

The successful application of ophthalmic VQA systems holds promise for benefiting both healthcare professionals and patients (**Figure 1b**).

**Ophthalmic Education**: Ophthalmic VQA systems serve as interactive learning tools for medical students and general practitioners, enhancing their understanding of ophthalmic diseases and management. These systems provide personalized answers based on diverse imaging, enabling learners to extract key points from ophthalmology textbooks and clinical guidelines. This facilitates exam revision and self-assessment for general practitioners.[47] While the interpretation of multimodal images is crucial in ophthalmology education, traditional training methods often provide limited exposure to typical cases. Consequently, medical students encountering real-world images may have numerous unanswered questions that ophthalmic specialists cannot promptly address. In such instances, students can upload their images and engage with the VQA system to access personalized learning opportunities.

**Clinical Assistant**: In real-world clinical practice, generating professional reports based on ophthalmic images is a highly specialized and time-consuming task,[48] particularly for junior doctors. VQA systems can assist junior doctors by providing accurate interpretations of ophthalmic images and generating automated preliminary reports. This assistance can lead to improved diagnostic accuracy, reduced diagnostic time, and faster decision-making processes. However, it's important to note that the outputs of the VQA system may still require modification by ophthalmologists to ensure accuracy and patient-specific relevance. Furthermore, the dynamic and interactive nature of the VQA system makes it a valuable resource for ophthalmologists in busy clinics, helping to minimize errors and enhance communication skills.

**Patient Consultant**: Ophthalmic VQA systems have the potential to revolutionize patient consultations by enabling remote triage and streamlining the consultation process. Firstly, integrating the VQA system into mobile devices or hospital websites can provide patients in regions with limited medical resources quick access to professional guidance and assistance in scheduling appointments at nearby hospitals. This approach promotes a more rational distribution of ophthalmic medical resources and mitigates the issue of overcrowding at tertiary ophthalmic centers.[49] Secondly, deploying VQA systems in outpatient consultations can aid patients by addressing their pre-consultation inquiries, identifying key questions to be discussed during face-to-face appointments, and providing clear explanations of relevant reports and prescriptions in plain language after the visit. This helps minimize the need for multiple physical queues and face-to-face interactions with doctors, thereby resulting in decreased outpatient burden, enhanced patient satisfaction, and reduced economic and time costs related to unnecessary follow-up.

**Utilizing LLM in Ophthalmic VQA**

LLMs, such as ChatGPT and Llama,[50 51] are trained on extensive datasets encompassing diverse topics, showcasing advanced capabilities in understanding, generating, and interacting with human language. These models have already shown potential applications in various medical contexts, including assisting in the generation of streamlining clinical documentation,[52 53] enhancing patient communication,[54] and answering ophthalmic questions[55]. By integrating LLMs into the framework of VQA, the analysis and interpretation of ophthalmic images can be significantly enhanced (**Figure 2**).

*LLMs as Data Creators*

In the field of ophthalmology, accurately annotated professional datasets are scarce but crucial for training effective VQA systems. Here, LLMs play a vital role as key data generators, enriching existing datasets or creating new ones by generating descriptive captions, synthetic patient inquiries, or detailed reports based on ophthalmic images. For instance, ophthalmic images and their corresponding reports, which are relatively easier to obtain, can be leveraged by LLMs to generate image-QA pairs, bridging the gap in dedicated VQA datasets.[37] Moreover, LLMs not only increase the quantity of training data for VQA models but also enhance their diversity, encompassing a wider range of eye diseases and scenarios. This is particularly beneficial for rare conditions like Stargardt disease and Leber hereditary optic neuropathy,[56] which are difficult to accumulate in the real world. The application of LLMs could facilitate the development of more robust and versatile VQA models.

*LLMs as Base Models*

Recently, general domain multimodal LLMs, such as LlaVA and GPT 4Vision,[28 57] which can process image data, have demonstrated commendable performance in VQA tasks on general images. However, their effectiveness when applied directly to ophthalmic images is suboptimal, with only 30.6% accuracy, 21.5% high usability, and 55.6% posing no harm.[11] Despite this limitation, LLMs, which have undergone extensive pretraining on a large scale of general images, possess a vast knowledge base and linguistic abilities, making them suitable as base models for further fine-tuning into specialized medical VQA systems. For example, models like LlaVA-Med have utilized large-scale biomedical image-text datasets extracted from PubMed Central. They have employed GPT-4 for prompt engineering and implemented novel fine-tuning methods to enhance the performance of the LLM LlaVA. This dedicated effort has resulted in the creation of a comprehensive language-vision assistant tailored for biomedicine.[58] Another notable model, Visual Med Alpaca, builds upon the LLaMA-7B architecture and is trained using diverse medical datasets sourced from the BigBIO repository.[59] Furthermore, Med PaLM M utilizes multimodal datasets comprising clinical texts, radiology and dermatology images, pathology slides, and genomics. It is built upon the PaLM-E model architecture and exhibits strong performance in various biomedical tasks.[60] Apart from fine-tuning, another approach to infusing ophthalmic expertise into LLMs is retrieval-augmented generation (RAG). For instance, ChatDoctor introduced online Wikipedia and manual datasets to enhance professional information integration, thus significantly improving the accuracy of responses.[61] Lewis et al. successfully combined pre-trained parameters with external knowledge for text generation.[62] Similarly, the RAG method was also explored to enhance the ophthalmic capabilities of Llama 2.[56] These explorations have yielded encouraging results in employing RAG technology to LLM-assisted VQA systems in ophthalmology.

*LLM as a part of the VQA pipeline*

LLMs can serve as a crucial part of the VQA pipeline, especially when it comes to optimizing free-form text outputs. This was demonstrated in FFA-GPT and ICGA-GPT systems.[12 46] These systems achieved this by combining LLMs with a fine-tuned image-text conversion model based on BLIP, enabling interactive QA after generating reports from FFA and ICGA images. This approach can enrich the output of the VQA pipeline by providing users with easily comprehensible language descriptions.

*LLMs as the Controllers*

LLMs can enhance VQA systems by acting like a human brain to schedule different modules to finish a task collaboratively. By combining LLMs with other AI models, these agents handle user requests and generate final answers.[63] When users ask questions, AI agents utilize LLMs for natural language understanding, along with image-text conversion models, to extract relevant information from images and texts and complete VQA tasks. This approach can enhance the precision and efficiency of vision-language information processing and leverage the strengths of diverse AI models, thereby enhancing the versatility and practicality of the VQA system.

The utilization of LLMs in ophthalmic VQA systems presents novel opportunities and prospects. Whether employed as data generators, foundational models, the central components of AI agents, or in other roles, LLMs pave the way for advancements in ophthalmic VQA systems. With continuous technological advancements and constant improvements in models, LLMs will gradually play an important role in enhancing the efficiency and intelligence of ophthalmic VQA systems in diagnosis and management.

**Challenges and Future Direction**

Despite the initial attempts at VQA in ophthalmology and the integration of LLMs into ophthalmic VQA systems, which have demonstrated great potential, we still face several challenges when it comes to real-world applications. Furthermore, the progress in this field has revealed several promising future directions that merit further exploration.

*VQA Datasets*

Ophthalmology, as a distinct branch of clinical medicine, encompasses a vast array of specialized terminology and imaging techniques. There is a critical need to establish high-quality datasets that specifically focus on ophthalmology and include annotations for training VQA models in this domain. There are two effective ways to establish these datasets: expanding existing ophthalmic image datasets and collecting real-world dialogues. Firstly, existing pure image datasets and multimodal image-text datasets can be utilized,[64] including classification,[65-67] segmentation,[68 69] and report generation datasets[70 71]. While these datasets may not explicitly provide QA pairs for training, they serve as valuable resources that can be leveraged to expand these pairs. For instance, by leveraging the excellent generation capabilities of LLMs, we can generate additional QA pairs from existing structured reports, thereby enriching the VQA dataset.[37] However, it is important to note that due to the potential for hallucinations generated by LLMs and it can be challenging to cover all possible scenarios. Therefore, it is important to also collect real dialogues from clinical practice to further enhance the dataset.

*Evaluation Methods*

Due to the unique nature of medicine, the evaluation specifically designed for medical applications of VQA is particularly important. The current evaluation methods for VQA systems mainly consist of automated assessment and human evaluation.[72] Automated assessment is a widely used method that employs several benchmarks, including MultiMedQA,[73] PubMedQA,[74] MedMCQA,[75] USMLE,[76] among others. The common evaluation metrics can be divided into two types: classification-based metrics, such as accuracy, recall, and F1 score; and language-based metrics, such as

BLEU,[77] BERTScore,[78] METEOR,[79] and ROUGE[80]. Automated assessments are widely used in evaluating existing medical VQA models due to their objectivity, computational automation, and simplicity. However, there is currently a lack of benchmark datasets for ophthalmology-related tasks, which means that most research can only utilize internal validation sets for evaluation. The medical field's demand for accurate and reliable answers necessitates the inclusion of evaluations by human doctors. Human evaluation is particularly suited for non-standard cases and open-ended tasks, such as medical consultations, where the quality and accuracy of model-generated results can extend beyond standard answers. For example, our team recently proposed a quantitative evaluation framework that involves ophthalmologists' evaluations of the model across accuracy, completeness, safety and satisfaction dimensions.[46] It is worth noting that no single metric can fully reflect the complexity of medical generation. Therefore, it is recommended to combine multiple methods and metrics to comprehensively assess the ophthalmic VQA system. Future research should focus on developing more precise indicators that reflect the comprehensive performance of ophthalmic VQA systems, encompassing not only accuracy and efficiency but also effectiveness in real clinical applications, including empathy, interpretability, and practical contribution to clinical processes.

*Clinical Application*

Although some initial ophthalmic VQA systems have demonstrated their capabilities in various scenarios, applying them to actual clinical environments requires addressing a series of practical issues. First, it is crucial to further optimize the VQA systems by fully leveraging the multimodal nature of ophthalmology. This involves exploring more effective ways to integrate and analyze different types of ophthalmic images and text data,[81] leading to more accurate disease diagnosis and management. Secondly, it is essential to improve the explainability of the VQA system. This will enable doctors and patients to understand the decision-making process of the model, thereby enhancing trust in the system's output.[82] Next, system integration and deployment need to be considered, particularly in relation to existing clinical workflows. For example, in eye disease screening, ophthalmic VQA combined with automatic imaging systems can be utilized for generating large-scale eye disease screening reports. Moreover, for telemedicine, ophthalmic VQA can serve patients in remote areas with limited medical resources. Finally, ethical considerations in the development of VQA-LLM are crucial.[83] Compliance with medical insurance regulations is paramount, and measures must be taken to ensure that these systems do not increase the workload on doctors. Furthermore, patient privacy and data security should be prioritized and carefully addressed.

By overcoming these challenges and exploring the aforementioned future directions, LLM-assisted ophthalmic VQA systems are expected to play a significant role in improving the accuracy of eye diagnoses, advancing medical education, enhancing patient experiences, and promoting global eye health.

**Conclusion**

VQA systems in ophthalmology, when enhanced with LLMs, have the potential to introduce a new era of precision, efficiency, and accessibility in medical practice and patient education. While developing specialized ophthalmology VQA systems poses challenges, such as the scarcity of VQA datasets and the complexities of adapting general AI models for medical-specific domains, the opportunities outweigh these barriers. We emphasize that widespread adoption of these systems in the medical field requires collaborative efforts from the global medical and AI research communities, coupled with rigorous clinical validation. By responsibly harnessing this potential, we envision a future where the obstacles to high-quality eye care are significantly reduced worldwide.

**Contributors**

DS developed the concept of the manuscript. XC, RC and PX drafted the manuscript. All authors have commented on the manuscript. DS and MH supervised the entire work. DS is the guarantor.

**Funding**

The study was supported by the Start-up Fund for RAPs under the Strategic Hiring Scheme (P0048623) from HKSAR, Global STEM Professorship Scheme (P0046113), and Henry G. Leong Endowed Professorship in Elderly Vision Health. The sponsors or funding organizations had no role in the design or conduct of this research.

**Acknowledgments**

We thank the InnoHK HKSAR Government for providing valuable supports.

**Competing Interests**

None declared.

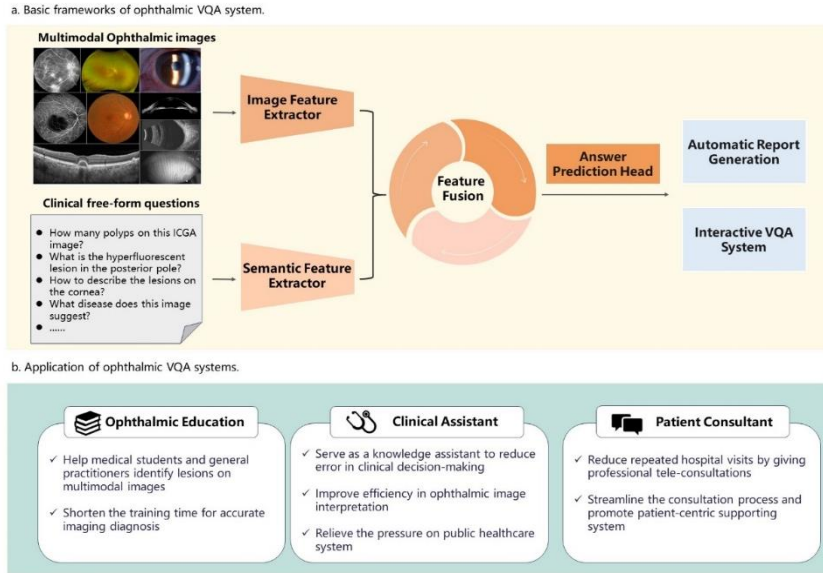

**Figure 1. The basic frameworks of ophthalmic VQA system and its applications**. VQA, Visual question answering.

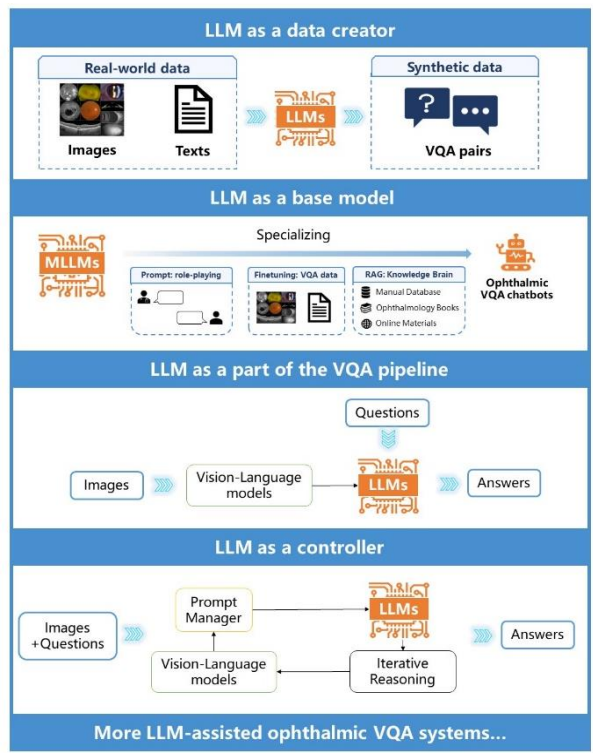

**Figure 2. Leverage the power of LLMs in VQA systems**. LLM, large language model; VQA, Visual question answering.